\documentclass[aps,pra,twocolumn,showpacs]{revtex4-1}

\usepackage{amsfonts,amssymb,amsmath}
\usepackage{mathtools}
\usepackage[]{graphics,graphicx,epsfig}
\usepackage{amsthm,multirow}
\usepackage{color}
\begin{document}

\title{Observable bound for Gaussian illumination}

\author{Su-Yong \surname{Lee}}\email[]{suyong2@add.re.kr}

\author{Yonggi Jo}\email[]{yonggi@add.re.kr}
\affiliation{Agency for Defense Development, Daejeon 34186, Korea}

\author{Taek Jeong}

\author{Junghyun Kim}

\author{Dong Hwan Kim}

\author{Dongkyu Kim}

\author{Duk Y. Kim}

\author{Yong Sup Ihn}

\author{Zaeill Kim}
\affiliation{Agency for Defense Development, Daejeon 34186, Korea}

\date{\today}

\begin{abstract}
We propose observable bounds for Gaussian illumination to maximize the signal-to-noise ratio, which minimizes the discrimination error between the presence and absence of a low-reflectivity target using Gaussian states. 
The observable bounds are achieved with mode-by-mode measurements. In the quantum regime using a two-mode squeezed vacuum state, 
our observable receiver outperforms the other feasible receivers whereas it cannot approach the quantum Chernoff bound (QCB). The corresponding observable cannot be implemented with heterodyne detections due to the additional vacuum noise.
In the classical regime using a thermal state, a receiver implemented with a photon number difference measurement approaches its bound regardless of the signal mean photon number, while it asymptotically approaches the classical bound in the limit of a huge idler mean photon number.
\end{abstract}

\maketitle

\section{Introduction}

Entanglement is a key element of quantum technologies, such as quantum teleportation, quantum communication, and quantum sensing. 
It takes advantage of the quantum correlation that cannot be revealed in classical systems.
Quantum illumination (QI), which belongs to quantum sensing, takes quantum advantage over classical illumination (CI), with no output entanglement\cite{Lloyd,Tan}.
QI is used to discriminate the presence and absence of a low-reflectivity target using entangled states that consist of signal and idler modes.
To detect the target, we send the signal mode towards the target while maintaining the idler mode. 
Then the reflected signal mode is measured together with the idler mode in a receiver. 
In continuous variable systems, a typical entangled state is a two-mode squeezed vacuum (TMSV) state which can be represented in terms of a number basis, $|\text{TMSV}\rangle=\sum^{\infty}_{n=0}\sqrt{\frac{N_S^n}{(1+N_S)^{n+1}}}|n\rangle_S|n\rangle_I$, where $N_S$ is the mean photon number of the signal (or idler) mode. The TMSV state is nearly optimal in QI\cite{Palma,Ranjith,Bradshaw}.
CI is used to detect the target using unentangled states, e.g., coherent or thermal states.
QI was compared with CI under a few measurement setups proposed in Refs.\cite{Guha,Sanz,Quntao,Karsa,Blakely, Lee, Jo, Jeffers} and implemented in Refs.\cite{Genovese,Zheshen,Sandbo,England,Aguilar,Luong,Shabir19,Sussman}.
To detect a long-distance target,  it was studied on microwave QI\cite{Shabir15} associated with the preparation of microwave signal and optical idler mode pairs. 
Even if optical entangled states are prepared, they can be converted to micro-optical entangled states by frequency conversion\cite{Lauk20,Ihn,Jo21}.

The performance of Gaussian illumination is quantified with the error probability that is a sum of the miss probability $P(\text{off}|\text{on})$ and false alarm probability $P(\text{on}|\text{off})$. Given a positive operator-valued measure,  the error probability is lower bounded by the Helstrom bound (HB) and upper bounded by the quantum Chernoff bound (QCB)\cite{Aud07,Calsa08,Stefano08} which is also upper bounded by the Bhattacharyya bound. 
It is not known how to achieve the HB with implementable setups, but the QCB can be achieved with feasible ones. A single-mode coherent state attains its QCB by homodyne detection, and a TMSV state approaches its QCB asymptotically by sum frequency generation with feedforward\cite{Quntao}.
Based on the QCB, QI using the TMSV state improves the error probability exponent by a factor of $4$ over CI using the coherent state\cite{Tan}.

Here, we consider a signal-to-noise ratio (SNR) under mode-by-mode measurements.
Initially we define that $\langle \hat{O}\rangle_i\equiv R_i$ is the mean value of an observable, $\Delta^2 O_i\equiv \langle \hat{O}^2\rangle_i-\langle \hat{O}\rangle^2_i$ is its variance, $i=\text{on}(1), \text{off}(0)$, and $M$ is the number of modes.
By repeated measurements on many copies $M\gg 1$, the sum of the measurements approaches a Gaussian distribution and subsequently it is applied to a decision threshold $R_{\text{th}}$.
 Given two Gaussian distributions with independent signal-idler mode pairs ($M\gg 1$), the total error probability is minimized according to the decision threshold. 
Each error probability is derived as follows: 
$P(\text{off}|\text{on})=\frac{1}{2}\text{erfc}[\frac{R_{\text{th}}-MR_1}{\sqrt{2M\Delta R_1}}]$, and $P(\text{on}|\text{off})=\frac{1}{2}\text{erfc}[\frac{MR_0 - R_{\text{th}}}{\sqrt{2M\Delta R_0}}]$. At $R_{\text{th}}=\frac{M(R_0\sqrt{\Delta R_1}+R_1\sqrt{\Delta R_0})}{\sqrt{\Delta R_0}+\sqrt{\Delta R_1}}$, the total error probability is minimized as  
$P^{(M)}_{\text{err}}=\frac{1}{2}[P(\text{off}|\text{on})+P(\text{on}|\text{off})]=\frac{1}{2}\text{erfc}[\sqrt{\text{SNR}^{(M)}}]$, where the complementary error function $\text{erfc}[z]=\frac{e^{-z^2}}{\sqrt{\pi}z}[1-\frac{1}{2(z^2+1)}+\frac{1}{4(z^2+1)(z^2+2)}-...]$. At $z\gg 1$, $\text{erfc}[z]\leq \frac{e^{-z^2}}{\sqrt{\pi}z} $ such that the minimum error probability is upper bounded by 
$ e^{-\text{SNR}^{(M)}}$, where $\sqrt{\pi\text{SNR}^{(M)}}$ is ignorable compared to $e^{\text{SNR}^{(M)}}$.
Thus, minimizing the error probability corresponds to maximizing the SNR which is explicitly given by
\begin{eqnarray}
\text{SNR}^{(M)}\equiv \frac{M(\langle \hat{O}\rangle_{\text{on}}-\langle \hat{O}\rangle_{\text{off}})^2}
{2(\sqrt{\Delta^2 O_{\text{on}}}+\sqrt{\Delta^2 O_{\text{off}}})^2}.
\label{SNR}
\end{eqnarray}
There are four known receivers, such as the phase conjugate (PC) receiver, the optical parametric amplifier (OPA) receiver\cite{Guha}, the double homodyne receiver\cite{Jo}, and the heterodyne receiver on each mode\cite{Sandbo}. We exclude the sum frequency generation with feedfoward\cite{Quntao}, which remains hard to implement due to its complicated structure requiring a sequence of nonlinear processes.


In this paper we propose an observable bound for Gaussian illumination, which maximizes the SNR with respect to mode-by-mode measurements.
In the quantum regime, we consider a TMSV state which  is described with a $4\times4$ covariance matrix $V_{\text{SI}}=\langle [\hat{a}_S~ \hat{a}_I~ \hat{a}_S^{\dag}~ \hat{a}_I^{\dag}]^T[ \hat{a}_S^{\dag}~ \hat{a}_I^{\dag}~ \hat{a}_S~ \hat{a}_I]\rangle$, where $S$ ($I$) represents the signal (idler) mode.
In the classical regime, an input two-mode state is prepared by impinging a coherent or thermal state into a beam splitter, which is described with the first moment and the covariance matrix. The input states interact with a target which is represented by a low-reflectivity beam splitter, where the thermal noise effect is simulated by impinging a thermal state into the beam splitter. Since both input states and interaction processes are in the Gaussian regime, we describe the output state with the covariance matrix and first moment.

\section{Observable bound for QI}
In QI, the signal mode is reflected from a target with reflectance $\kappa$ while the idler mode is kept ideally. The output covariance matrix\cite{Guha} that represents target-on is given by
\begin{align}
	 V_{\text{SI}}(\kappa)=\begin{pmatrix}
 A+1 & 0 & 0 & C \\
0 & N_S+1 & C & 0 \\
0 & C & A & 0\\
C & 0 &0 & N_S
	\end{pmatrix},
\end{align}
where $A=\kappa N_S+N_B$, $C=\sqrt{\kappa N_S(N_S+1)}$, and $N_B$ is the mean photon number of thermal noise. 
When the target is off,  the covariance matrix becomes $V_{\text{SI}}(0)$.
 The covariance matrix $V_{\text{SI}}(\kappa)$ differs from the covariance matrix $V_{\text{SI}}(0)$ by the correlation elements
$(\langle\hat{a}^{\dag}_S\hat{a}^{\dag}_I + \hat{a}_S\hat{a}_I\rangle$) and the mean photon number of the signal mode 
($\langle\hat{a}^{\dag}_S\hat{a}_S\rangle$).
Thus, we propose an observable with a combination of the three elements and the mean photon number of the idler mode $\langle\hat{a}^{\dag}_I\hat{a}_I\rangle$,
\begin{eqnarray}
\hat{O}_{\text{prs}}=\hat{a}^{\dag}_S\hat{a}^{\dag}_I+\hat{a}_S\hat{a}_I+\alpha\hat{a}^{\dag}_S\hat{a}_S+\beta\hat{a}^{\dag}_I\hat{a}_I ,
\label{prs}
\end{eqnarray}
where $\alpha$ and $\beta$ are real values.
The idler mode element is not known for any effect on the performance, and
the other elements of the covariance matrix are useless due to only increasing the variance in the SNR.
Previously, the PC receiver\cite{Guha} measures the observable $\hat{O}_{\text{PC}}=\nu(\hat{a}^{\dag}_S\hat{a}^{\dag}_I+\hat{a}_S\hat{a}_I)+\mu(\hat{a}_I\hat{a}^{\dag}_V+\hat{a}_V\hat{a}^{\dag}_I)$, where $|\mu|^2-|\nu|^2=1$ and $\hat{a}_V$ is a vacuum state operator.  
The OPA receiver\cite{Guha} measures the observable $\hat{O}_{\text{OPA}}=\sqrt{G(G-1)}(\hat{a}^{\dag}_S\hat{a}^{\dag}_I+\hat{a}_S\hat{a}_I)
+(G-1)\hat{a}_S\hat{a}^{\dag}_S+G\hat{a}^{\dag}_I\hat{a}_I$, where $G>1$ is a gain of the OPA.
The double homodyne (DH) receiver\cite{Jo} measures the observable  $\hat{O}_{\text{DH}}=(\hat{X}^2_S+\hat{P}^2_I)=-(\hat{a}^{\dag}_S\hat{a}^{\dag}_I+\hat{a}_S\hat{a}_I)+\hat{a}_S\hat{a}^{\dag}_S+\hat{a}^{\dag}_I\hat{a}_I$.
 Since the correlation elements had a fixed relation with the other elements in the previous receivers, it is worthwhile to investigate a general relation between the correlation elements and the other ones. Note that the OPA and double homodyne receivers include the vacuum noise as $\langle\hat{a}_S\hat{a}^{\dag}_S\rangle=1+\langle\hat{a}^{\dag}_S\hat{a}_S\rangle$ but Eq. (3) does not.

Here, we consider QI under not only constant thermal noise but also non constant thermal noise that corresponds to the passive signature case \cite{Stefano18}.
In a table-top experiment, it is natural to think about the non constant thermal noise since a thermal noise is injected into a beam splitter that represents a target. 
When we put the initial mean value of the thermal noise being constant without any relation with the beam-splitting parameter, the output thermal noise is related to the beam splitting parameter after the interaction with the beam-splitter, resulting in nonconstant thermal noise.

\subsection{constant thermal noise}
Equation (3) is optimized by maximizing the SNR of Eq.~(\ref{SNR}), such that we call it the bound observable,
\begin{eqnarray}
\hat{O}_{\text{bd}}=\hat{a}^{\dag}_S\hat{a}^{\dag}_I+\hat{a}_S\hat{a}_I-|\beta|\hat{a}^{\dag}_I\hat{a}_I ,
\end{eqnarray}
and the SNR
\begin{eqnarray}
\text{SNR}^{(M)}_{\text{bd}}= \frac{2MC^2}
{\bigg[\sqrt{\Delta^2 O_{\kappa}+m(\kappa)}+\sqrt{\Delta^2 O_{0}+m(0)}\bigg]^2},\nonumber\\
\label{optSNR}
\end{eqnarray}
where $\Delta^2 O_{\kappa}=(A+1)(N_S+1)+2C^2+AN_S,~m(\kappa)=|\beta|^2N_S(N_S+1)-2|\beta|C(2N_S+1).$
$\Delta^2 O_{\kappa}$ is a variance of a nearly bound observable, $\hat{O}_{\text{SI}}=\hat{a}^{\dag}_S\hat{a}^{\dag}_I+\hat{a}_S\hat{a}_I$, when a target is on.
Since the element $\langle\hat{a}^{\dag}_S\hat{a}_S\rangle$ measures the mean photon number which is constituted with a small amount of the reflected signal and a large amount of the transmitted thermal noise, it increases the variance dominantly rather than the mean value.
However, the element  $\langle\hat{a}^{\dag}_I\hat{a}_I\rangle $ measures only the idler mode so that it can help reduce the variance. Thus, the value of $\alpha$ goes to zero and the $\beta$ survives with variance contribution $m(\kappa)$. 
We present an analytic formula of $|\beta|$ with a plot in Appendix A.
When $N_S$ is much smaller than $N_B$, 
we can also ignore the $|\beta|$, resulting in a nearly bound observable as $\hat{O}_{\text{SI}}=\hat{a}^{\dag}_S\hat{a}^{\dag}_I+\hat{a}_S\hat{a}_I$ which measures only the off-diagonal elements. By taking the nearly bound observable, we obtain the SNR,
\begin{eqnarray}
\text{SNR}^{(M)}_{\text{nbd}}= \frac{2MC^2}
{\bigg[\sqrt{\Delta^2 O_{\kappa}}+\sqrt{\Delta^2 O_{0}}\bigg]^2}.
\label{nSNR}
\end{eqnarray}

\begin{figure}
\includegraphics[width=0.45\textwidth]{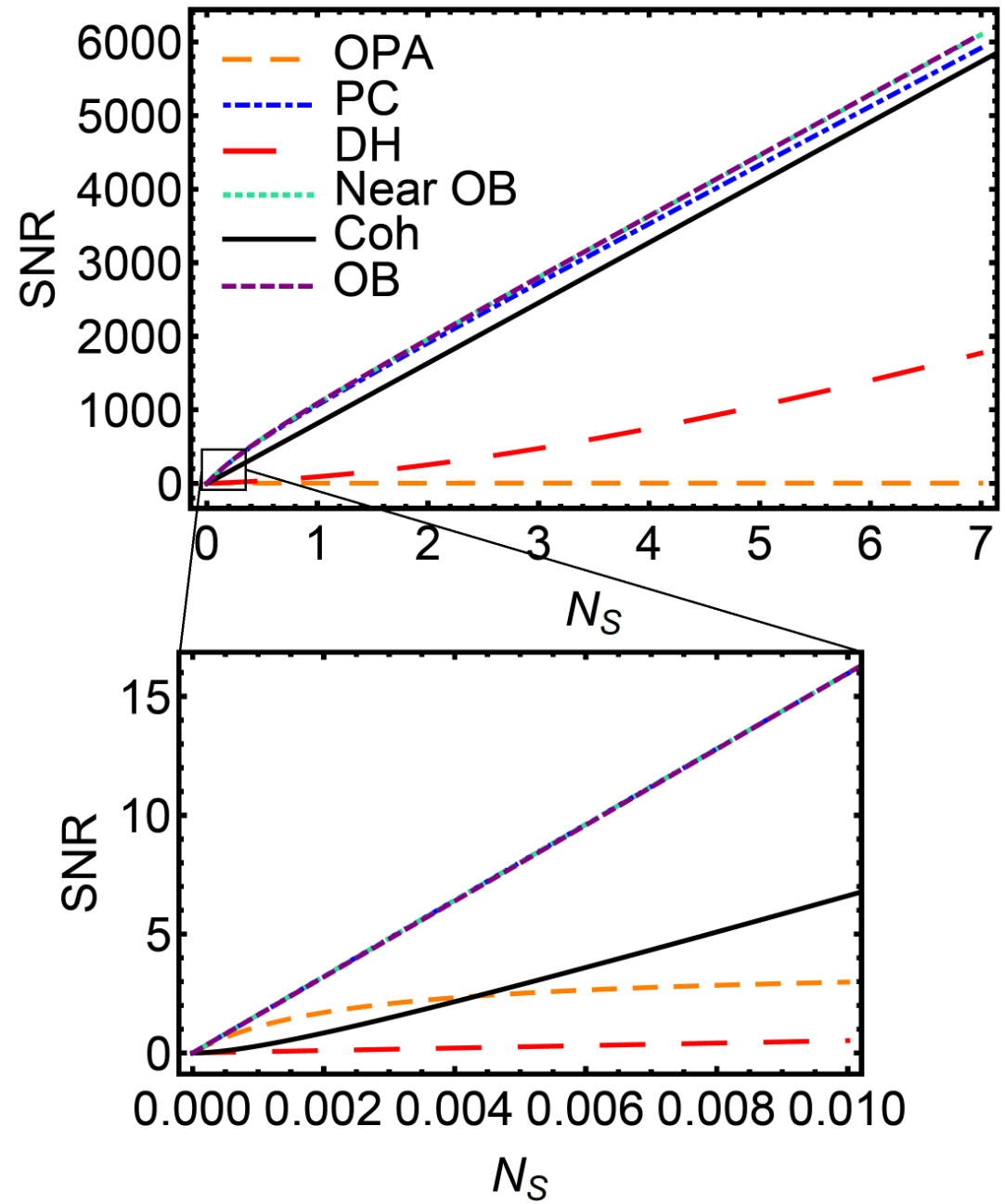}
\caption{SNR for QI as a function of $N_S$ at $\kappa=0.01$, $N_B=30$, $M=10^7$ under constant thermal noise: coherent-state bound (Coh), receivers with observable bound (OB), nearly observable bound (OB), PC, OPA, and double homodyne (DH).  
}
\label{fig:fig1}
\end{figure}

\begin{figure}
\includegraphics[width=0.47\textwidth]{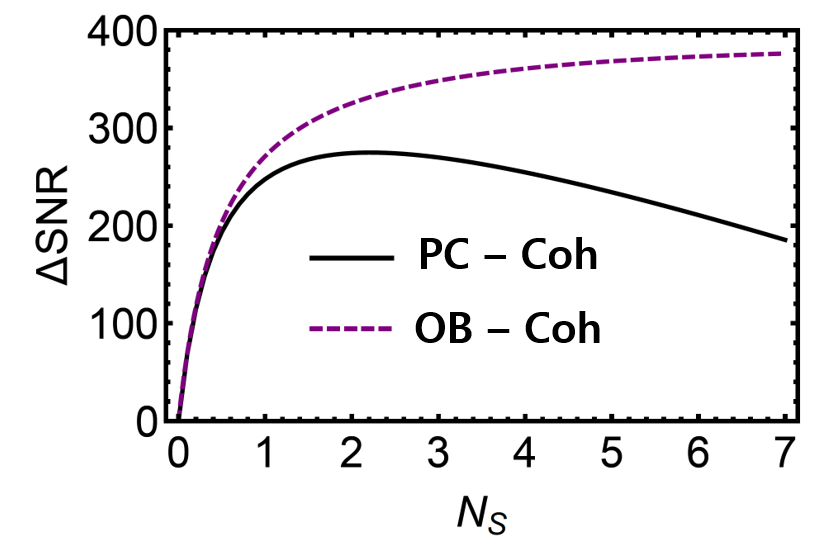}
\caption{ SNR difference as a function of $N_S$ at $\kappa=0.01$, $N_B=30$, $M=10^7$ under constant thermal noise: between the bound observable receiver and coherent-state bound (purple dashed curve), and between the PC receiver and the coherent-state bound (black solid curve). 
}
\label{fig:fig2}
\end{figure}

In Fig. \ref{fig:fig1}, we observe that the SNR of Eq.~(\ref{nSNR}) is overlapped with the SNR of Eq.~(\ref{optSNR}).
Quantitatively, the bound observable receiver exhibits approximately $3\%$ higher SNR than the nearly bound observable receiver.
It outperforms the SNRs of other observable receivers, such as PC, OPA, and double homodyne receivers, while beating the coherent-state QCB. Analytic formulas of the other receivers are given in Appendix A.
At a low input power $N_S<0.01$, the PC receiver is also overlapped with the bound and nearly bound observable receivers, as shown in the inset of Fig. \ref{fig:fig1}.
 In the limit of $N_S,\kappa\ll 1$ and $N_B\gg 1$, the $\text{SNR}^{(M)}_{\text{SI}}$ asymptotically approaches 
$\frac{M\kappa N_S}{2N_B}$ and its error probability becomes 
$P^{(M)}_{\text{err}}=\frac{1}{2}\text{erfc}[\sqrt{\frac{M\kappa N_S}{2N_B}}]\leq e^{-\frac{M\kappa N_S}{2N_B}}$.
With increasing $N_S$, however, the SNR difference between the bound observable receiver and the coherent-state bound increases by more than twice the SNR difference between the PC receiver and the coherent-state bound, as shown in Fig.\ref{fig:fig2}. At $N_S=7$, the SNR difference for the bound observable receiver is about $376$ whereas the SNR difference for the PC receiver is about $185$.

\begin{figure}
\includegraphics[width=0.43\textwidth]{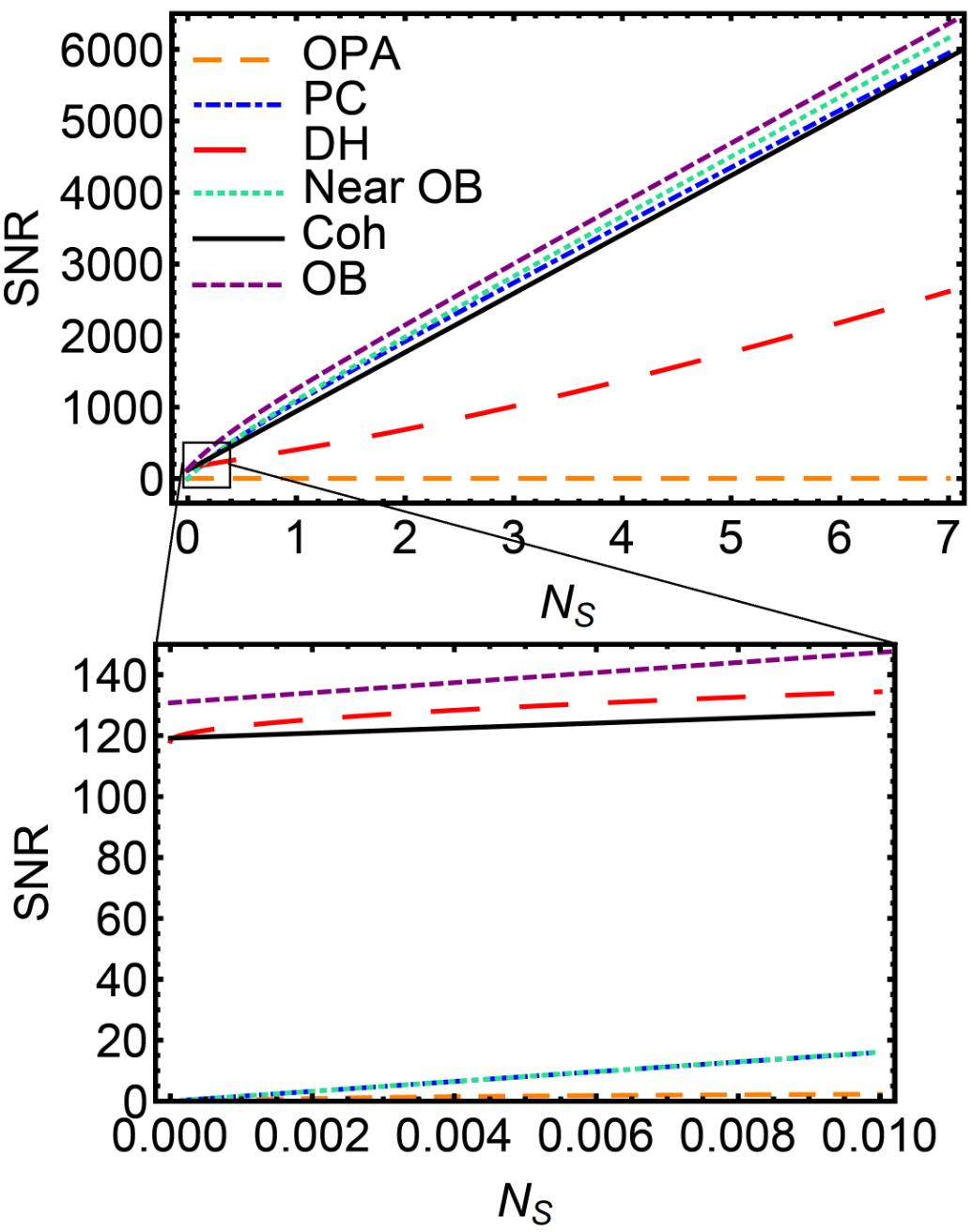}
\caption{SNR for QI as a function of $N_S$ at $\kappa=0.01$, $N_B=30$, $M=10^7$ under nonconstant thermal noise: coherent-state bound, receivers with observable bound (OB), nearly observable bound (OB), PC, OPA, and double homodyne (DH). 
}
\label{fig:fig3}
\end{figure}

\subsection{nonconstant thermal noise}
In the above, we simulated the thermal noise $N_B$ by injecting a thermal state with mean photon number $N_B/(1-\kappa)$ into a beam splitter with reflectance $\kappa$, leading to the thermal noise that is independent of a target reflectance. We can raise a question regarding if we designate the thermal mean photon number as not 
$N_B/(1-\kappa)$ but $N_B$ which is natural in experiments. Thus, the thermal contribution depends on the target reflectance as $(1-\kappa)N_B$.
Since the thermal noise contains the target information as $\kappa N_B$ which is not ignorable,  therefore, the bound observable should include
the last component $\langle\hat{a}^{\dag}_S\hat{a}_S\rangle$ of Eq.~(\ref{prs}).
The corresponding SNR is given by
\begin{eqnarray}
\text{SNR}^{(M)}_{\text{bd,n}}= \frac{M\bigg[2C-\alpha\kappa(N_B-N_S)\bigg]^2}
{2\bigg[\sqrt{\Delta^2 O_{\kappa,n}+l(\kappa)}+\sqrt{\Delta^2 O_{0,n}+l(0)}\bigg]^2},\nonumber\\
\label{optSNRn}
\end{eqnarray}
where $\Delta^2 O_{\kappa,n}=2C^2+BN_S+(B+1)(N_S+1),
~l(\kappa)=\alpha^2B(B+1)+\beta^2N_S(N_S+1)+2\alpha C(2B+1)+2\beta C(2N_S+1)+2\alpha\beta C^2$, 
and $B=\kappa N_S+(1-\kappa)N_B$. 
$\Delta^2 O_{\kappa,n}$ is a variance of the observable, $\hat{O}_{\text{SI}}=\hat{a}^{\dag}_S\hat{a}^{\dag}_I+\hat{a}_S\hat{a}_I$, when a target is on.
 We numerically obtain negative values of $\alpha$ and $\beta$ whose details are plotted in Appendix B.
In Fig. \ref{fig:fig3}, we compare all the receivers under nonconstant thermal noise.  The SNR of Eq.~(\ref{optSNRn}) outperforms all the other receivers. At a low input power $N_S<0.01$, the double homodyne receiver is close to the bound one, as shown in the inset of Fig. \ref{fig:fig3}.
In the limit of $N_S\rightarrow 0$, the SNR of Eq.~(\ref{optSNRn}) converges to a non zero value because the transmitted thermal noise contains the target information that is measured with $\langle\hat{a}^{\dag}_S\hat{a}_S\rangle$. This also applies to the results of the double homodyne receiver and the coherent-state bound.

\subsection{Measurement setup}
\begin{figure}
\includegraphics[width=0.45\textwidth]{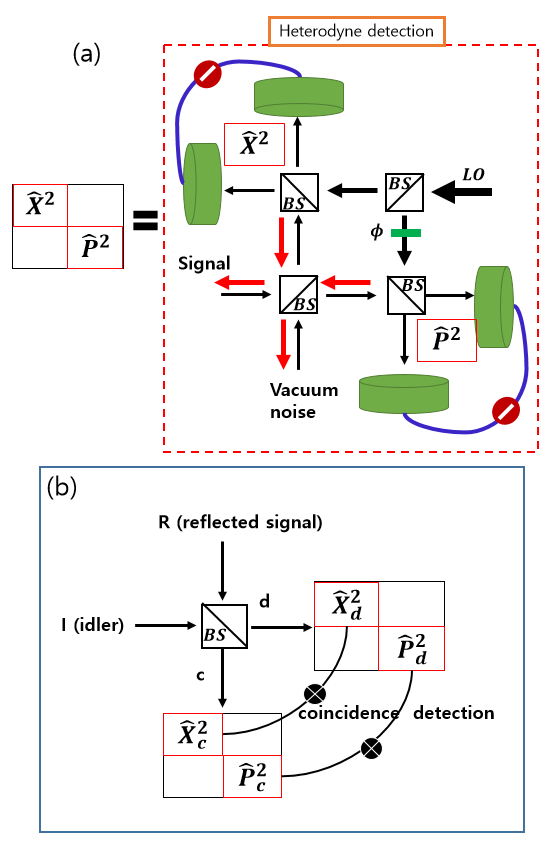}
\caption{Measurement setups for QI: (a) HTD, and (b) coincidence HTDs after combining the reflected and idler modes.
BS represents a $50:50$ beam splitter. $\phi$ is a phase shifter. LO means a local oscillator.
When we perform the HTD, the reverse arrow (red color) indicates how to derive an observable including vacuum noise.
It includes the following observables, $\hat{X}_c\hat{X}_d$, $\hat{P}_c\hat{P}_d$, $\hat{X}^2_{c(d)}$, and $\hat{P}^2_{c(d)}$. }

\label{fig:S3}
\end{figure}

In Fig. \ref{fig:S3}, we consider if it is possible to implement the bound observable as well as the nearly bound one, in terms of linear optics and heterodyne detection (HTD)
that performs homodyne detection (HD) on each mode after dividing a signal by a $50:50$ beam splitter.
HD measures a quadrature operator as follows:
A signal and a local oscillator (LO) are impinged on a $50:50$ beam splitter, and then we measure the intensity difference between the output ports.
Thus, we obtain the mean value of a quadrature operator as
$\langle \hat{n}_a-\hat{n}_b\rangle=|\alpha_L|\langle \hat{X}(\phi)\rangle=|\alpha_L|\langle \frac{\hat{a}^{\dag}e^{i\phi}+\hat{a}e^{-i\phi}}{\sqrt{2}}\rangle$.
The phase (or amplitude) of the LO controls $\phi$ (or $\alpha_L$). At $\phi=0$ (or $\pi/2$), the quadrature operator corresponds to the position $\hat{X}$ (or momentum $\hat{P}$) operator. Both position and momentum operators on the signal can be measured by the HTD with additional vacuum noise.


\begin{figure}
\includegraphics[width=0.45\textwidth]{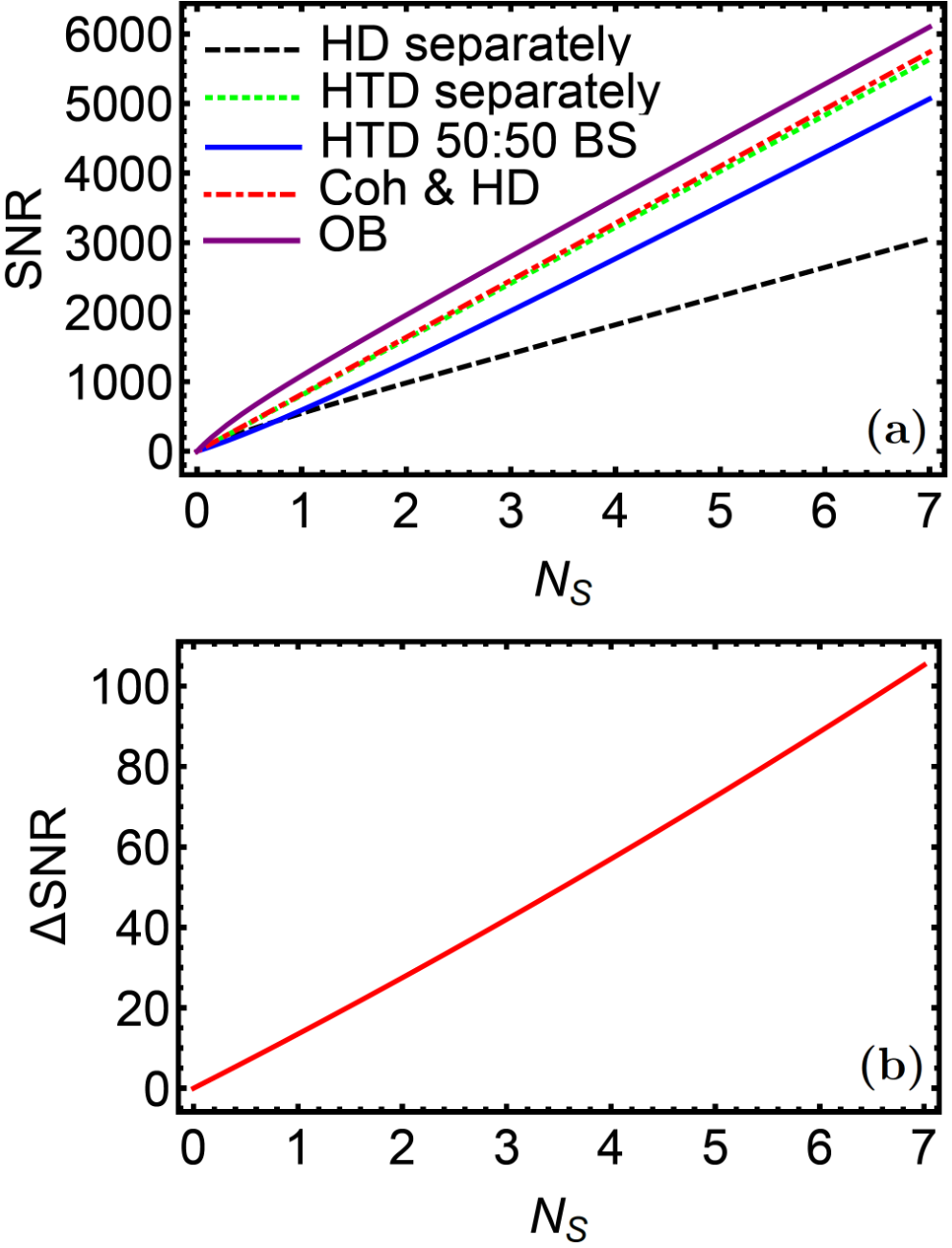}
\caption{QI bound by dHTD under constant thermal noise: (a) SNR as a function of $N_S$ at $\kappa=0.01$, $N_B=30$, $M=10^7$,
 and (b) SNR difference between the coherent state with HD and TMSV state with HTD separately. The coherent state outperforms the TMSV state under the HD and HTD. Coh \& HD represents a coherent state with homodyne detection. OB represents observable bound.
}
\label{fig:fig4}
\end{figure}

However, in the double HTD (dHTD) after the $50:50$ beam splitter, the additional vacuum noise produces a large variance so that we cannot take quantum advantage over the classical bound. Let us consider the case of the nearly bound observable $\hat{O}_{\text{SI}}=\hat{a}^{\dag}_S\hat{a}^{\dag}_I+\hat{a}_S\hat{a}_I$ which is almost overlapped with the case of the bound observable.
In Fig. \ref{fig:fig4}(a), we show that the TMSV state with double HTD cannot beat the performance of the coherent state with HD because additional vacuum noise produces more errors than the information that we get by the double HTD. If we directly measure the observable $\hat{O}_{\text{SI}}=\hat{a}^{\dag}_S\hat{a}^{\dag}_I+\hat{a}_S\hat{a}_I$ by performing HTD on the reflected and idler modes separately, the performance is still below the performance of the coherent state with HD, as shown in Fig.\ref{fig:fig4} (b). 
Moreover, if we directly measure an observable $\hat{O}_{\text{HD}}=\hat{X}_R(\theta)\hat{X}_I(\phi)$ (optimized at $\theta+\phi=n\pi$, $n=0,1,2,...$) by  performing HD on the reflected and idler modes separately, the performance is getting worse than the performance by the double HTD after the $50:50$ beam splitter
with increasing $N_S$. More information is given in Appendix C.

\section{Observable bound for CI}
\begin{figure}
\includegraphics[width=0.45\textwidth]{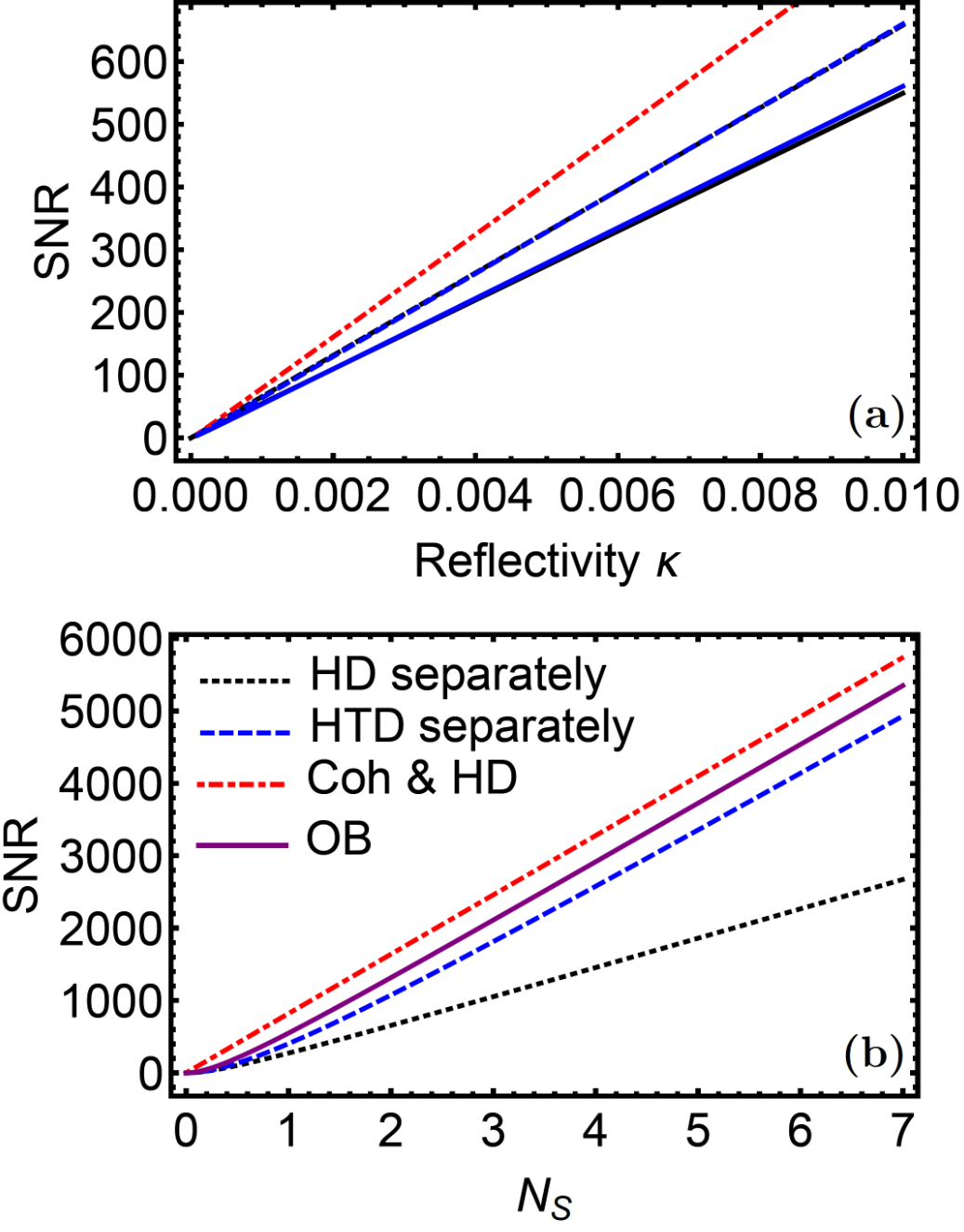}
\caption{For $N_B=30,~M=10^7$, (a) SNR for CI as a function of target reflectance $\kappa$ ;
CCT state by QCB (blue) and by $\hat{O}_{\text{off}}$ (black) at $N_S=N_I=1$ (solid lines) and $N_S=1, N_I=2$ (dashed lines).
(b) SNR for CI as a function of $N_S$ at $\kappa=0.01$ ; CCT states at $N_S=N_I$ by QCB (blue curve) and $\hat{O}_{\text{off}}$ (black curve) that are overlapped. Coherent state by QCB is described with the red dotted-dashed lines.
}
\label{fig:fig5}
\end{figure}

In CI, a single-mode coherent state asymptotically attains its QCB by performing homodyne detection\cite{Tan},
but it is not known if a classically correlated thermal state can approach its QCB under mode-by-mode measurements and how far it is from the coherent-state QCB.
Instead of a single-mode input state, we consider a two-mode input state which is produced by impinging  a thermal state into a beam splitter,
resulting in a classically correlated thermal (CCT) state.  
The output covariance matrix that represents target-on is given by
\begin{align}
	 C_{\text{SI}}(\kappa)=\begin{pmatrix}
 B+1 & D & 0 & 0 \\
D & N_I+1 & 0 & 0 \\
0 & 0 & B & D\\
0 & 0 &D & N_I
	\end{pmatrix},
\end{align}
where $B=\kappa N_S+N_B,~D=\sqrt{\kappa N_SN_I}$. $N_S~(N_I)$ is the mean photon number of the signal (idler) mode that is controlled by the beam-splitting ratio. The off-diagonal element $D$ produces a classical correlation. 
When the target is off, the covariance matrix becomes $C_{\text{SI}}(0)$.
By comparing the covariance matrices of the target-on and target-off, 
we propose an observable $\hat{O}_{\text{off}}=\hat{a}^{\dag}_S\hat{a}_I+\hat{a}^{\dag}_I\hat{a}_S$, consisting of off-diagonal elements.
The corresponding SNR is given by
\begin{eqnarray}
\text{SNR}^{(M)}_{\text{th}}= \frac{2M\kappa N_SN_I}
{\bigg[\sqrt{4\kappa N_SN_I+\kappa N_S+y}+\sqrt{y}\bigg]^2},
\label{cSNR}
\end{eqnarray}
where $y\equiv N_I+N_B(1+2N_I)$. The SNR with the observable attains the QCB, as shown in Fig.\ref{fig:fig5}. 
Under a fixed $N_S$, the amount of the classical correlation is proportional to $N_I$, resulting in an enhanced SNR. However it cannot beat the coherent-state QCB.
Replacing the thermal state by a coherent state, 
we also obtain $\text{SNR}^{(M)}_{\text{coh}}$ by removing the term $4\kappa N_SN_I$ in the denominator of Eq.~(\ref{cSNR}). But the coherent state with the observable cannot attain its QCB.
In Fig. \ref{fig:fig5}, the coherent state with HD shows a higher SNR than the thermal state. 
The tendencies are maintained regardless of constant or nonconstant thermal noise.
In the limit of $N_S,\kappa\ll 1$ and $N_B, N_I\gg 1$, the $\text{SNR}^{(M)}_{\text{th}}$ asymptotically approaches 
$\frac{M\kappa N_S}{4N_B}$, and its error probability becomes 
$P^{(M)}_{\text{err}}=\frac{1}{2}\text{erfc}[\sqrt{\frac{M\kappa N_S}{4N_B}}]\leq e^{-\frac{M\kappa N_S}{4N_B}}$, 
resulting in a coherent-state QCB.

\subsection{Measurement setup}
The off-diagonal elements represent the bound observable. It is implemented by a photon number difference measurement (PNDM) after interfering the reflected and idler modes by a $50:50$ beam splitter. The PNDM observable is described with $\hat{c}^{\dag}\hat{c}-\hat{d}^{\dag}\hat{d}$, and it is inversely transformed to $\hat{a}^{\dag}_S\hat{a}_I+\hat{a}^{\dag}_I\hat{a}_S$ by the $50:50$ beam splitter. Given a general beam-splitting transformation of $\hat{c}^{\dag}\rightarrow t\hat{a}^{\dag}_S-ie^{-i\varphi}r\hat{a}^{\dag}_I$ and $\hat{d}^{\dag}\rightarrow t\hat{a}^{\dag}_I-ie^{i\varphi}r\hat{a}^{\dag}_S$, the transformed observable is derived as
\begin{eqnarray}
&&\hat{c}^{\dag}\hat{c}-\hat{d}^{\dag}\hat{d} \nonumber\\
&&\rightarrow
(t^2-r^2)(\hat{a}^{\dag}_S\hat{a}_S-\hat{a}^{\dag}_I\hat{a}_I) \\
&&-2rt[(\hat{a}^{\dag}_S\hat{a}_I+\hat{a}^{\dag}_I\hat{a}_S)\sin\varphi
-i(\hat{a}^{\dag}_S\hat{a}_I-\hat{a}^{\dag}_I\hat{a}_S)\cos\varphi ].\nonumber
\label{PNDM}
\end{eqnarray}
At $t=r=\frac{1}{\sqrt{2}}$ and $\varphi=(2n+1)\frac{\pi}{2}~(n=0,1,2,...)$, the transformed observable becomes 
$\pm(\hat{a}^{\dag}_S\hat{a}_I+\hat{a}^{\dag}_I\hat{a}_S)$.
The phase component $\varphi$ includes a phase shifter in one arm that plays the role of a path-length difference.
In the limit of $N_I\gg 1$, the PNDM converges to homodyne detection as
$\langle \hat{a}^{\dag}_S\hat{a}_I+\hat{a}^{\dag}_I\hat{a}_S \rangle \rightarrow \sqrt{2 N_I}\langle \hat{X}_S(\phi)\rangle$.
It is the same as CI using a single-mode coherent state with homodyne detection.

Moreover, we may consider the observable $\hat{O}_{\text{off}}=\hat{a}^{\dag}_S\hat{a}_I+\hat{a}^{\dag}_I\hat{a}_S=\hat{X}_S\hat{X}_I +\hat{P}_S\hat{P}_I$ by performing HTD on the reflected and idler modes separately.  
After including the vacuum noise, the corresponding mean and variance are given by
\begin{eqnarray}
\langle \hat{O}_{\text{off}, v}\rangle&=&\frac{1}{2}\langle \hat{O}_{\text{off}}\rangle,\\
\Delta^2 \hat{O}_{\text{off}, v}&=&
\frac{1}{4}[\Delta^2 \hat{O}_{\text{off}}+2+\langle \hat{n}_S+\hat{n}_I\rangle].\nonumber
\end{eqnarray}
We obtain that the performance of the HTD is below the performance of the bound observable.  If we consider an observable 
$\hat{O}_{HD}=\hat{X}_R(\theta)\hat{X}_I(\phi)$ (optimized at $\phi-\theta=n\pi$, $n=0,1,2,...$) by performing HD on the reflected and idler modes separately, the performance is even worse than the performance of the HTD.

\section{Discussion}
In Gaussian illumination, we proposed bound observables with feasible measurement setups to maximize the SNR under constant and nonconstant thermal noises. Using the bound observables under the condition of $N_S,\kappa\ll 1$ and $N_B\gg 1$, we showed that QI using entangled states cannot attain its QCB but CI using classically correlated states can attain its QCB.  In QI, the SNR using the bound observable outperforms the SNRs using other observables for any number of $N_S,~N_B,~\kappa$. 
However, the bound observable measurement cannot be achieved by using linear optics with heterodyne detection due to the additional vacuum noise. 
In CI, the measurement setup consists of PNDM after combining the reflected and idler modes by a $50:50$ beam splitter. 
The SNR using the bound observable can asymptotically approach the coherent-state QCB, while the SNR with a classically correlated thermal state cannot beat the SNR with a coherent state.

Since the observables we considered do not include collective measurements, the receivers with the observables do not always approach the QCBs.
Note that, in the limit of $N_S\ll 1$, QI using our receiver asymptotically improves the error probability exponent by a factor of $2$ over the classical state QCB.
It guarantees a half exponent of the TMSV state QCB by measuring the elements of the output covariance matrix. 
Since our bound observables belong to a class of local operations assisted with classical communication\cite{Calsa10,Bandy11}, it cannot approach the QCB that requires collective protocols\cite{Quntao}. However, CI using our receiver asymptotically approaches the classical bound in the limit of $N_S \ll1$ and 
$N_I\gg1$.

In QI, we could not measure the bound observable with linear optics, HD, and HTD. 
As a further work, it remains an open question of how the bound observable can be  experimentally measured with nonlinear systems.


\begin{acknowledgments}
This work was supported by a grant to Defense-Specialized Project funded by Defense Acquisition Program Administration and Agency for Defense Development.

\end{acknowledgments}

\section*{appendix A: SNR under constant thermal noise}
\begin{figure}
\includegraphics[width=0.42\textwidth]{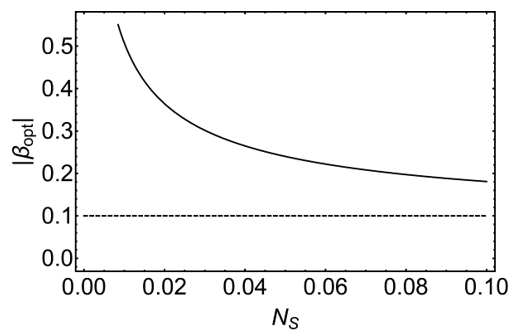}
\caption{Optimal value of $|\beta |$ of Eq.~(5) as a function of $N_S$ at $\kappa=0.01$, $N_B=30$, $M=10^7$ under constant thermal noise.}
\label{fig:S1}
\end{figure}

For the SNR of Eq.~(5) using the bound observable, we derive the corresponding optimal relation $|\beta|=\frac{(1+2N_S)}{\sqrt{\kappa N_S(N_S+1)^3}}[f-\sqrt{f(f-\kappa(N_S+1))}]$, where $f=1+N_S+N_B+2N_SN_B$. 
It is shown in Fig. \ref{fig:S1}, which corresponds to the values of $|\beta|$ in Fig.1. 
With increasing $N_S$, the value of $|\beta|$ converges to $\sqrt{\kappa}$.

From the observables considered in Sec. II, we analytically derive the SNRs of three different receivers under constant thermal noise after the target interaction.
For the PC receiver, the SNR is given by
\begin{eqnarray}
\text{SNR}^{(M)}_{\text{PC}}= \frac{2MC^2}
{\bigg[\sqrt{\Delta^2 O_{\kappa}+\frac{\mu^2}{\nu^2}N_S}+\sqrt{\Delta^2 O_{0}+\frac{\mu^2}{\nu^2}N_S}\bigg]^2}, \nonumber\\
\label{PC}
\end{eqnarray}
which is always smaller than the SNR of Eq.~(6). The parameters are designated as $\mu=\sqrt{2}$ and $\nu=1$\cite{Guha}. 

For the OPA receiver, the SNR is given by
\begin{eqnarray}
\text{SNR}^{(M)}_{\text{OPA}}= \frac{2M\bigg(C+\sqrt{\frac{G-1}{G}}\frac{\kappa N_S}{2}\bigg)^2}
{\bigg[\sqrt{\Delta^2 O_{\kappa}+q(\kappa)}+\sqrt{\Delta^2 O_{0}+q(0)}\bigg]^2}, \nonumber\\
\label{OPA}
\end{eqnarray}
where $q(\kappa)=\frac{G-1}{G}A(A+1)+\frac{G}{G-1}N_S(N_S+1)+\frac{C}{\sqrt{G(G-1)}}[(G-1)(4A+2)+G(4N_S+1)]+2C^2$. 
The gain is implementable as $G-1=7.4\times 10^{-5}$\cite{Zheshen}.
Since the additional denominator terms $q(\kappa)$ and $q(0)$ are much larger than the additional numerator term $\sqrt{\frac{G-1}{G}}\frac{\kappa N_S}{2}$, 
the SNR with the OPA is always smaller than the SNR of Eq.~(6).

For the double homodyne (DH) receiver, the SNR is given by
\begin{eqnarray}
\text{SNR}^{(M)}_{\text{DH}}= \frac{2M\bigg(C-\frac{\kappa N_S}{2}\bigg)^2}
{\bigg[\sqrt{\Delta^2 O_{\kappa}+p(\kappa)}+\sqrt{\Delta^2 O_{0}+p(0)}\bigg]^2}, \nonumber\\
\label{DH}
\end{eqnarray}
where $p(\kappa)=A(A+1)+N_S(N_S+1)+2C^2-4C(A+N_S+1)$.
Since all the additional terms diminish the SNR, the SNR with the DH is always smaller than the SNR of Eq.~(6).\\

\begin{figure}
\includegraphics[width=0.42\textwidth]{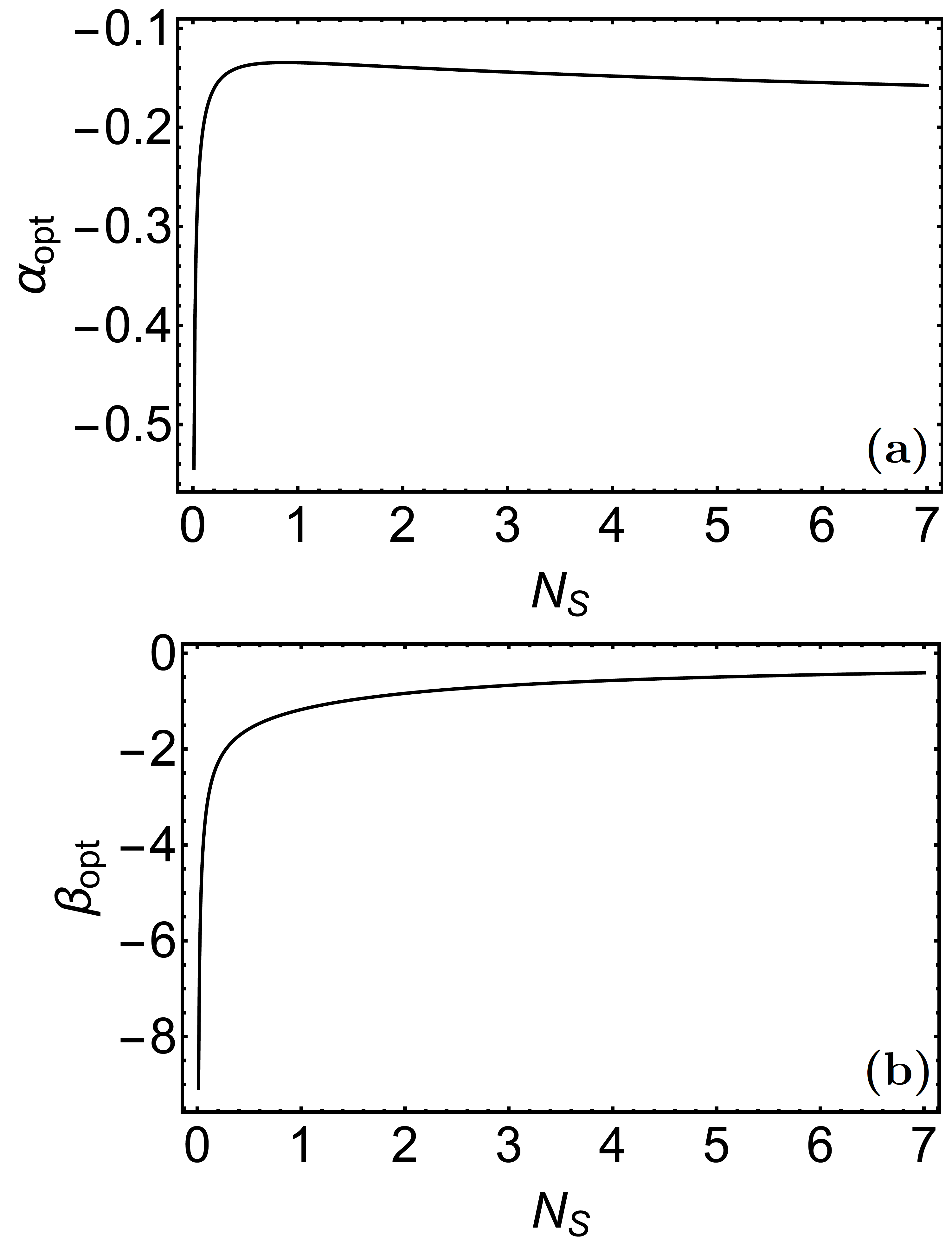}
\caption{Optimal values of $\alpha$ and $\beta$ of Eq.~(7) as a function of $N_S$ at $\kappa=0.01$, $N_B=30$, $M=10^7$ under non constant thermal noise.}
\label{fig:S2}
\end{figure}

\section*{appendix B: SNR under nonconstant thermal noise}
Initially, we assign the thermal mean photon number as $N_B$ which is natural in experiments.
After the target interaction, the transmitted thermal noise becomes $(1-\kappa)N_B$, and then the matrix element $A$  is transformed as $\kappa N_S+N_B\rightarrow\kappa N_S+(1-\kappa)N_B$. For the SNR of Eq.~(7) using the bound observable, we obtain the values of $\alpha$ and $\beta$ numerically, 
as shown in Fig. \ref{fig:S2}.
At $N_S=0.01$, the $\alpha$ and $\beta$ correspond to $-0.54$ and $-9.08$, respectively.
In Eqs.~(\ref{OPA}) and (\ref{DH}),  due to the observable shapes, a component of the numerator is transformed as $\kappa N_S/2\rightarrow\kappa(N_S-N_B)/2$.

\section*{appendix C: Measurement setup using double heterodyne detection}
After combining the reflected and idler modes by a $50:50$ beam splitter, $\hat{a}^{\dag}_S\rightarrow \frac{1}{\sqrt{2}}(\hat{c}^{\dag}+\hat{d}^{\dag})$ and $\hat{a}^{\dag}_I\rightarrow \frac{1}{\sqrt{2}}(\hat{d}^{\dag}-\hat{c}^{\dag})$, we obtain the output observable
\begin{eqnarray}
\hat{M}_{prs}(\alpha,\beta)
&=&\frac{1}{2}\bigg[(\frac{\alpha+\beta}{2}+1)(\hat{X}^2_d+\hat{P}^2_c)\nonumber\\
&&+(\frac{\alpha+\beta}{2}-1)(\hat{X}^2_c+\hat{P}^2_d)-(\alpha+\beta)\nonumber\\
&&+(\alpha-\beta)(\hat{X}_c\hat{X}_d+\hat{P}_c\hat{P}_d)\bigg],
\label{prsob}
\end{eqnarray}
which demands coincidence measurements on the position and momentum operators, together with the square of each operator. 
The mean value of a square quadrature operator is obtained by calculating the squared outcomes of HD, 
$\langle \hat{X}^2(\phi)\rangle=\int^{\infty}_{-\infty}dx x^2P(x,\phi)$, where the marginal distribution $P(x,\phi)$ is obtained by repeated measurements.

However it is unavoidable to produce additional vacuum noise after the HTD.
Let us see the observable
$\hat{M}_{prs}(0,0)=\frac{1}{2}[(\hat{X}^2_d-\hat{P}^2_d)-(\hat{X}^2_c-\hat{P}^2_c)]$ that corresponds to the nearly bound observable. 
As shown in Fig. \ref{fig:S3} (a), an additional vacuum noise is included into the observable by the transformation of 
$\hat{X}_{d(c)}\rightarrow \frac{1}{\sqrt{2}}(\hat{X}_{d(c)}+\hat{X}_{d(c),v})$ and
$\hat{P}_{d(c)}\rightarrow \frac{1}{\sqrt{2}}(\hat{P}_{d(c)}-\hat{P}_{d(c),v})$, where 
$\hat{X}_{d(c),v}$ and $\hat{P}_{d(c),v}$ belong to the vacuum noise.
Although we measure $\hat{X}^2_{d(c)}$ and $\hat{P}^2_{d(c)}$ by heterodyne detection on the output signal mode, 
due to the vacuum noise, the observable is transformed into
\begin{widetext}
\begin{eqnarray}
\hat{M}_{prs, v}(0,0)=\frac{1}{2}[\hat{M}_{prs}(0,0)+\frac{1}{2}(\hat{X}^2_{d,v}-\hat{P}^2_{d,v})
-\frac{1}{2}(\hat{X}^2_{c,v}-\hat{P}^2_{c,v})+(\hat{X}_d\hat{X}_{d,v}+\hat{P}_d\hat{P}_{d,v})
+(\hat{X}_c\hat{X}_{c,v}+\hat{P}_c\hat{P}_{c,v})],
\label{vacuum}
\end{eqnarray}
\end{widetext}
which is applied to the output $c$ and $d$ modes, as shown in Fig.\ref{fig:S3} (b).
Then, the mean and variance of Eq. (\ref{vacuum}) are given by
\begin{eqnarray}
\langle \hat{M}_{prs, v}(0,0)\rangle&=&\frac{1}{2}\langle \hat{M}_{prs}(0,0)\rangle,\\
\Delta^2 \hat{M}_{prs, v}(0,0)&=&
\frac{1}{4}[\Delta^2 \hat{M}_{prs}(0,0)+1+\langle \hat{n}_c+\hat{n}_d\rangle].\nonumber
\end{eqnarray}
The output variance of Eq.~(\ref{nSNR}) is enlarged as $\Delta^2 O_{\kappa}+4[1+N_B+(1+\kappa)N_S]$, which is close to the output variance of the bound observable including the vacuum noise.

On the other hand, we directly measure the observable $\hat{O}_{\text{SI}}=\hat{a}^{\dag}_S\hat{a}^{\dag}_I+\hat{a}_S\hat{a}_I$ by performing HTD on the reflected and idler modes separately. After including the vacuum noise, the corresponding mean and variance are given by
\begin{eqnarray}
\langle \hat{O}_{\text{SI}, v}\rangle&=&\frac{1}{2}\langle \hat{O}_{\text{SI}}\rangle,\\
\Delta^2 \hat{O}_{\text{SI}, v}&=&
\frac{1}{4}[\Delta^2 \hat{O}_{\text{SI}}+1+\langle \hat{n}_S+\hat{n}_I\rangle].\nonumber
\end{eqnarray}
The corresponding output variance is $\Delta^2 O_{\kappa}+[1+N_B+(1+\kappa)N_S]$.

\end{document}